\documentclass[a4paper,10pt]{article}
\usepackage[english]{babel} 
\usepackage{multicol}
\usepackage{amsfonts,mathrsfs,amssymb}
\usepackage[colorlinks]{hyperref}
\usepackage{flushend}
\usepackage{balance}
\usepackage{amsmath}
\usepackage{authblk}
\usepackage{xcolor}

\newcommand{\pa}{\partial}
\newcommand{\de}{\mathrm{d}}

\newcommand{\bx}{\mathbf{x}}
\newcommand{\Tr}{\operatorname{Tr}}

\newcommand{\rem}[1]{}

\usepackage[left=0.5in, right=0.5in, top=0.5in, bottom=2cm]{geometry}
\renewenvironment{abstract}{
\hspace{\fill}
\begin{minipage}{0.85\textwidth}
}
{
\end{minipage}\hspace{\fill}}

\newtheorem{theorem}{Theorem}

\newtheorem{lemma}[theorem]{Lemma}

\numberwithin{theorem}{section}

\begin{document}
\vspace{-2cm}
\title{Equivalent {\color{ black}variational} approaches\\to biaxial liquid crystal dynamics }
\date{}
\author{Alexander R.D. Close, Cesare Tronci}
\affil{\small\it Department of Mathematics, University of Surrey, Guildford GU2 XH, United Kingdom}
\maketitle
\begin{abstract}
Within the framework of liquid crystal flows, the Qian \& Sheng (QS) model \cite{QS1998} for $Q-$tensor dynamics is compared to the Volovik \& Kats (VK) theory \cite{VK1981} of biaxial nematics by using Hamilton's variational principle. Under the assumption of rotational dynamics for the $Q-$tensor, the {\color{ black} variational principles underling the two theories are equivalent and the conservative VK theory emerges as a specialization of the QS model.} Also, after presenting a micropolar variant of the VK model, Rayleigh dissipation is included in the treatment. Finally, the treatment is extended to account for nontrivial eigenvalue dynamics in the VK model and this is done by considering the effect of scaling factors in the evolution of the $Q-$tensor.
\end{abstract}

\vspace{1cm}
\begin{multicols}{2}

\section{Introduction}

Following the celebrated Ericksen-Leslie (EL) model \cite{ericksen1991liquid,Leslie1979} of uniaxial liquid crystals, several dynamical theories have been formulated over the decades. These theories often differ by the choice of order parameter (depending on the phase under consideration) and even different director formulations are available in the simplest case of uniaxial molecules. For example, while the EL theory defines the director as an unsigned (position dependent) unit vector in physical space, the Harvard theory \cite{forster1971hydrodynamics} defines the same order parameter as a differential one-form or vector field, as elucidated in Volovik's work \cite{volovik1980relationship}. This difference leads to intrinsically distinct transformation properties of the director field, which are reflected in substantial differences between the EL and the Harvard theories.

In the description of other nematic phases, such as the biaxial phase, de Gennes \cite{deGennes1969} showed the convenience of identifying the order parameter with a traceless symmetric tensor, the $Q-$tensor. This has the advantage of incorporating features of different liquid crystal phases (e.g. uniaxial and biaxial) by simply writing different expressions of the same tensor order parameter in terms of the director fields. Again, different theories of $Q-$tensor dynamics are available. Other than by the expression of the free energy or dissipation function, some of these theories again differ by intrinsic transformation properties of the $Q-$tensor. For example, in \cite{stark2003poisson} the alignment tensor is a (degenerate) covariant tensor field, while in Qian \& Sheng \cite{QS1998} the same tensor is simply a (position dependent) $3\times 3$ matrix. As before, this leads to substantial differences between the theories. In other situations, alignment tensor theories differ in whether or not inertial effects are retained in the rotational dynamics. When these effects are neglected, dissipation dominates to drive relaxation dynamics. For example, inertial rotation terms are neglected in the $Q-$tensor dynamics of Volovik \& Kats \cite{VK1981}, while the theory of Beris \& Edwards \cite{delaware1994thermodynamics} neglects inertial effects to formulate $Q-$tensor dynamics as a gradient flow on the space of symmetric covariant tensor fields. More generally, dissipation can be introduced by either the use of symmetric brackets \cite{delaware1994thermodynamics} (which are added to the Poisson bracket governing inertial effects) or the use of Rayleigh dissipation in Hamilton's variational principle \cite{sonnet2004continuum, Edwards2005131}.

In recent years, different approaches to director dynamics were shown to be equivalent when the director is defined as a unit vector. Indeed, while it is well known how the EL theory can also be formulated in terms of the molecular angular velocity (see Volovik's works \cite{Volovik1980,VK1981}), new developments \cite{gay2013equivalent,GBRaTr2012} have shown how several liquid crystal theories are actually special cases of Eringen's theory of micropolar fluids \cite{Eringen1993,Eringen1997}. Based on the variational approach, variants of these theories are constructed depending on how one expresses the free energy and the dissipation function. {\color{ black}As a result, these studies allowed the identification of the hydrodynamic helicity invariant for the conservative limit of several liquid crystal dynamical models \cite{gay2010helicity}.}

In this paper, we compare the $Q-$tensor dynamical theories of Qian \& Sheng (QS) \cite{QS1998} and Volovik \& Kats (VK) \cite{VK1981}. More particularly, upon restoring inertial effects in the VK theory (see also \cite{gay2010reduction}), we show that {\color{ black}its dissipationless limit arises as a specialization of the QS model}, under the common assumption of rotational dynamics for {\color{ black}biaxial liquid crystal phases}, that is \cite{GKSS2013}
\begin{equation}\label{evolution}
Q=RQ_0R^T
\end{equation}
(where $R(\mathbf{x},t)$ is a rotation matrix).
This evolution naturally arises from de Gennes' definition (see e.g. \cite{rey2012rheological})
\begin{equation}\label{Qdef}
Q=\int_{S^2}\!\left(\boldsymbol{u}\boldsymbol{u}-\frac13\boldsymbol{1}\right)f(\boldsymbol{u})\,\de^2\boldsymbol{u}\,,
\end{equation}
where $f$ is the statistical distribution of the molecular unit vector $\boldsymbol{u}$ (pointing along the long molecular axis) on the unit sphere $S^2$.
We shall show that using the above evolution in the variational principle for the QS model returns VK theory augmented to allow for inertial effects. More particularly, upon following Edwards' approach in \cite{Edwards2005131}, most of this paper focuses on the use of Hamilton's principle to formulate conservative dynamics, and this picture is eventually extended by the insertion of Rayleigh dissipation \cite{sonnet2004continuum}.
After discussing the variational principle for the conservative limit of the QS model for liquid crystal textures, Section 2 makes use of the evolution \eqref{evolution} to recover the VK theory in the absence of fluid flow. In addition, Eringen's wryness tensor is introduced as an auxiliary variable to obtain a micropolar variant of the VK model. Then, Section 3 extends the treatment to flowing liquid crystals. Section 4 deals with Rayleigh dissipation upon making use of the evolution \eqref{evolution} eliminating the need for the constraint $\operatorname{Tr}Q=0$ appearing in QS theory. Finally, Section 5 extends the treatment to consider nontrivial eigenvalue dynamics for the $Q-$tensor and this is done upon invoking the existence of a scaling factor, by mimicking the properties of microstretch fluids \cite{eringen2001microcontinuum1,eringen2001microcontinuum2,gay2009geometric}.

\section{Conservative texture dynamics: QS and VK theories}

In this section, we show that the QS model {\color{ black}specializes} to the VK theory, under the assumption of rotational evolution \eqref{evolution}. Upon considering the special case of a flowless nematic texture in the absence of dissipation, we shall derive the QS and VK theories from Hamilton's variational principle, along the lines of Edwards \cite{Edwards2005131}. In particular, we shall make use of the rotational symmetry by applying standard Euler-Poincar\'e techniques in geometric mechanics, following the works of Holm \cite{holm2002euler} and Gay-Balmaz \& Ratiu \cite{gay2009geometric}.

Upon denoting by $J$ the moment of inertia density, the conservative QS model for  $Q-$tensor dynamics in a nematic texture reads
\begin{equation}\label{QStexture}
	J\pa_t^2{Q}-h
	=\lambda_0 \boldsymbol{1}
\end{equation}
where $h$ is the molecular field
\[
h=-\frac{\pa \mathcal{F}}{\pa Q}+\operatorname{div}\!\left(\frac{\pa \mathcal{F}}{\pa\nabla Q}\right)
\,,
\]
$\mathcal{F}$ is the Landau-de Gennes free energy
\begin{multline*}
	\mathcal{F}=\frac{1}{2}(\alpha Q^2_{ij}+L_1Q^2_{ij,k}+L_2Q_{ij,j}Q_{ik,k})
	\\\qquad
	+\frac{4\pi}{P}L_1\varepsilon_{ijk}Q_{il}\pa_jQ_{kl}-\beta Q_{ij}Q_{jk}Q_{ki}
	+\gamma(Q^2_{ij})^2
\end{multline*}
with $\alpha,L_1,L_2,\beta,\gamma,P$ phenomenological constants,
and $\lambda_0$ is a Lagrange multiplier enforcing $\Tr Q=0$, so that taking the trace of \eqref{QStexture} yields
\[
\lambda_0=\frac13\Tr\!\left(J\partial^2_t{Q}-h\right)=-\frac13\Tr h
\,.
\]
In the original paper \cite{QS1998}, an extra Lagrange multiplier enforces $Q=Q^T$, although here we drop the corresponding term by suitably symmetrising the molecular field such that $h=h^T$. The QS equation \eqref{QStexture} is evidently an Euler-Lagrange equation on the space of symmetric matrices, arising from the constrained variational principle
\begin{equation}
\delta\!\int_{t_1}^{t_2}\!\!\!\int\! \left(\frac{J}{2}\|\dot Q\|^2
-\mathcal{F}(Q,\nabla Q)
	+\lambda_0\Tr Q\right)\de^3{x}\,\de t=0\,,
	\label{QStextActPrinc}
\end{equation}
where the last term is the constraint enforcing the traceless condition. Recently, a similar approach was  followed for biaxial {\color{ black}phases} in \cite{GKSS2013}, where constrained variational principles were used to prescribe the values of the first three rotational invariants  in the sequence $\{\Tr Q^n\}$. On the other hand, in the same work, it is recognized how biaxial molecules require the $Q-$tensor to undergo purely rotational evolution of the type \eqref{evolution}. In again the same paper, the authors emphasize how the change of position of a molecule in space can be described by the angular velocity matrix, which in our notation reads
\[
\widehat{\nu}=(\partial_t{R})R^{-1}
\]
and whose corresponding angular velocity vector is $\nu_i=-\epsilon_{ikj}\widehat{\nu}_{jk}/2$. It is the purpose of this paper to introduce this variable explicitly in the QS model and show how the resulting equations coincide with the VK model in \cite{VK1981}, when the latter is augmented to consider inertial effects arising from $J\neq0$. {\color{ black}An immediate verification of this statement is obtained upon noticing that \eqref{evolution} implies ${\pa_t Q}=[\widehat{\nu},Q]$, which is then replaced in \eqref{QStexture}: then, taking the commutator of the latter equation with $Q$ yields the conservative VK equation \cite{VK1981} $J{\pa_t}[[\widehat{\nu},Q],Q]=\big[h,Q\big]$. The remainder of this paper shows that the reason for the appearance of these equations lies in the fact that the QS and VK models share the same Hamilton's variational principle, under the ansatz \eqref{evolution}. Therefore, their underlying minimization problems are equivalent.}

In order to introduce the variable $\widehat{\nu}$ {\color{ black}in the variational framework}, we apply Euler-Poincar\'e theory \cite{holm1998euler,holm2002euler,gay2009geometric} by replacing the evolution relation \eqref{evolution} in the action principle \eqref{QStextActPrinc}. This operation yields the Euler-Poincar\'e variational principle
\[
\delta\!\int_{t_1}^{t_2}\!\!\!\int\! \left(\frac{J}{2}\|[\widehat{\nu},Q]\|^2d^3x
-\mathcal{F}(Q,\nabla Q)
	\right)\de^3{x}\,\de t=0\,,
\]
where the constraint has now been dropped appropriately and the rotational invariants $\Tr Q^n$ no longer contribute to the Landau-de-Gennes free energy. Then, upon computing the Euler-Poincar\'e variations
\begin{equation}\label{vars1}
\delta\widehat{\nu}=\partial_t{\widehat{\eta}}+[\widehat{\eta},\widehat{\nu}]
\,,\qquad\quad
\delta Q=[\widehat{\eta},Q]
\,,
\end{equation}
with $\widehat{\eta}=\delta{R} R^{-1}$, one is led to the Euler-Poincar\'e equations
\begin{equation}\label{VKdyn}
	J\frac{\pa}{\pa t}\big[[\widehat{\nu},Q],Q\big]
	=\big[h,Q\big]
	\,,\qquad\ \ 
	\frac{\pa Q}{\pa t}=[\widehat{\nu},Q]
	\,.
\end{equation}
which coincide with the conservative limit of the VK model in \cite{VK1981}, upon setting $J=0$ to neglect inertial effects (and in the absence of fluid flow). These effects are also recovered in the VK model by retaining the corresponding terms in the Poisson bracket structures appearing in \cite{VK1981}. Notice that the traceless condition has never been imposed in this treatment because it naturally arises from a convenient choice of initial condition $\Tr Q_0 =0$. {\color{ black}This Euler-Poincar\'e variational approach was recently followed in \cite{BlGoKa} to study textures of biaxial nematic phases in an external field.}

Perhaps not surprisingly, the above equations combine into
\[
\big[J\pa_t^2{Q}-h,Q\big]=0\,,
\]
which is solved{\color{ black}, for example,} by
\[
	J\pa_t^2Q-h
	=\lambda_nQ^n
	\,,
\]
for arbitrary time dependent functions $\lambda_n$. Alternatively, if $Q$ had not been considered \emph{a priori} as obeying \eqref{evolution}, then  $\lambda_n$ would correspond to a sequence of Lagrange multipliers prescribing the values of $\Tr Q^{n+1}$. As an example, in \cite{GKSS2013}, the first three $\lambda_n$'s were considered, while the QS theory  uses $\lambda_0$ only. 
{\color{ black}Consequently, the VK model emerges as a more reliable alternative to the approach followed in \cite{GKSS2013}, since the rotational dynamics is intrinsically encoded in the model. Indeed, the Euler-Poincar\'e formulation of the VK model was recently adopted in \cite{BlGoKa}.}

In conclusion, we have proved that the conservative VK theory {\color{ black}emerges from the QS model, under} the assumption \eqref{evolution} of purely rotational dynamics of biaxial molecules. This relation between the two theories arises from the fact that both  are derived from exactly the same action principle \eqref{QStextActPrinc}.

A micropolar theory of the above equations can also be obtained by noticing that \cite{gay2013equivalent}
\begin{align*}
	\partial_i Q
	&=[\partial_i RR^{-1}, Q]+R\partial_i Q_0R^{-1}
	\\
	&=[\partial_i RR^{-1}, Q]+R[Q_0,\widehat{\gamma}_{0i}]R^{-1}
	\\
	&=[Q,-\partial_i RR^{-1}+R\,\widehat{\gamma}_{0i} R^{-1}]
	\\
	&=:[ Q,\widehat{\gamma}_i]
\end{align*}
where we have  invoked the existence of an initial \emph{wryness tensor}  such that $\partial_i Q_0=[Q_0,\widehat{\gamma}_{0i}]$. In Eringen's micropolar theory \cite{Eringen1997}, the time-dependent wryness tensor
\[
\widehat{\gamma}_i:=-\partial_i RR^{-1}+R\,\widehat{\gamma}_{0i} R^{-1}
\]
identifies the amount of rotation under an infinitesimal displacement $\de\mathbf{x}$ and thus determines the spatial rotational strain \cite{holm1998euler}. Upon computing 
\[
\delta\widehat{\gamma}_i
	=[\widehat{\eta},\widehat{\gamma}_i]-\partial_i\widehat{\eta}
\]
and by writing $\mathcal{F}(Q,[Q,\widehat{\gamma}])=\Psi(Q,\widehat\gamma)$, one obtains the equations of motion
\begin{align}\label{ErText1}
		&J\frac{\pa}{\pa t}\big[[\widehat{\nu},Q],Q\big]
	=\left[Q,\frac{\partial\Psi}{\partial Q}\right]
	+\left[\frac{\partial\Psi}{\partial\widehat{\gamma}_i},\widehat{\gamma}_i\right]
	-\partial_i\frac{\partial\Psi}{\partial\widehat{\gamma}_i},
	\\ 
\label{ErText2}
	&\frac{\pa Q}{\pa t}=[\widehat{\nu},Q]
\,,\qquad \qquad 
\frac{\pa \widehat{\gamma}_i}{\pa t}
	=[\widehat{\nu},\widehat{\gamma}_i]-\partial_i\widehat{\nu}.
\end{align}
These equations represent the micropolar version of the VK model. In the gauge theory of defects, an inhomogeneous initial condition on the wryness tensor can be associated with the presence of disclinations in the texture \cite{Volovick1980,holm1998euler,GBRaTr2012}. 

Now that we have characterized how the QS and VK theories are related in the case of conservative texture dynamics, and have provided a micropolar variant of the VK model of biaxial liquid crystals, we shall proceed to consider the more general case of flowing liquid crystals. In this case the rotational degrees of freedom are coupled to the relabelling properties that characterize fluid flows.

\section{QS and VK theories for liquid crystal flows\label{sec:flows}}
As in the previous section, we start by focusing on the following QS equations for liquid crystal dynamics:
\begin{align}\label{QS1}
&\rho D_t \boldsymbol{u}= -\nabla
p-\partial_l\!\left(\frac{\partial
\mathcal{F}}{\partial{Q}_{ij\,,l}}\nabla{Q}_{ij}\right),
\\ \label{QS2}
&\rho JD_t^2Q-h=\lambda_0\boldsymbol{1}
\\ \label{QS3}
&
D_t \rho+\rho\operatorname{div}\boldsymbol{u}=0
\end{align}
where $D_t=\partial_t+\boldsymbol{u}\cdot\nabla$ is the ordinary convective derivative. Notice that the above equations are a slight generalization of the original QS equations in \cite{QS1998}, as they allow for a compressible fluid flow and the pressure is expressed in terms of an internal energy function $\mathcal{U}(\rho)$ as $p=\rho^2\mathcal{U}'-\mathcal{F}$. It is straightforward to observe that equations \eqref{QS1}-\eqref{QS2} arise from the following Eulerian action principle:
\begin{multline}
\delta\!\int_{t_1}^{t_2}\!\!\!\int\! \bigg(\frac12\rho\|\boldsymbol{u}\|^2+\frac{J}{2}\rho\|D_t Q\|^2
-\rho\,\mathcal{U}(\rho)
\\
-\mathcal{F}(Q,\nabla Q)
	+\lambda_0\Tr Q\bigg)\,\de^3{x}\,\de t=0\,,
	\label{QStextActPrincFluid}
\end{multline}
with variations
\begin{align*}
\delta\boldsymbol{u}&=\partial_t\boldsymbol{w}+(\boldsymbol{u}\cdot\nabla)\boldsymbol{w}-\boldsymbol{w}\cdot\nabla\boldsymbol{u}
\\
\delta\rho&=-\operatorname{div}(\rho\boldsymbol{w})
\\
\delta Q&=-(\boldsymbol{w}\cdot\nabla)Q+\Theta
\\
\delta (D_tQ)&=-(\boldsymbol{w}\cdot\nabla)D_tQ+D_t\Theta
\end{align*}
for arbitrary $\boldsymbol{w}$ and $\Theta$ vanishing at the endpoints.

Although the form of the variations above may look somewhat mysterious, they find a natural justification in terms of the relabelling properties of the action in \eqref{QStextActPrincFluid}.  Here we shall follow an analogous treatment to that in \cite{GBRaTr2012}.
In order to show how this works, we shall introduce the Lagrangian fluid path $\mathbf{X}(\mathbf{x}_0,t)$ and its corresponding Lagrange-to-Euler map
\[
\rho(\mathbf{x},t)=\int\!\rho_0(\mathbf{x}_0)\,\delta(\mathbf{x}-\mathbf{X}(\mathbf{x}_0,t))\,\de^3 x_0=:\mathbf{X}_*\rho_0
\,,
\]
whose fundamental role is expressing the Eulerian density $\rho(\mathbf{x}, t)$ in terms of its (fixed) Lagrangian correspondent $\rho_0(\mathbf{x}_0)$. Then following Euler-Poincar\'e theory we write the Eulerian velocity as $\boldsymbol{u}(\mathbf{x},t)=\dot{\mathbf{X}}(\mathbf{X}^{-1}(\mathbf{x},t),t)$ and observe that the relabelling property of the action in \eqref{QStextActPrincFluid} takes it into the form:
\begin{multline}
\label{ActPrinc2}
\int_{t_1}^{t_2}\!\!\!
\int\! \bigg(\frac12\rho_0\|\dot{\mathbf{X}}\|^2+\frac{J}{2}\rho_0\|\dot{\mathcal{Q}}\|^2
-\rho_0\,\mathcal{U}(\mathbf{X}_*\rho_0\circ\mathbf{X})
\\
-\frac{\rho_0}{(\mathbf{X}_*\rho_0)\circ\mathbf{X}}\,\mathcal{F}(\mathcal{Q},\nabla \mathcal{Q})
	+(\mathbf{X}_*\lambda_0)\Tr \mathcal{Q}\bigg)\,\de^3{x}\,\de t
\end{multline}
where we have introduced $\mathcal{Q}=Q\circ\mathbf{X}$ and $\circ$ denotes standard composition of functions.
In turn, the above action identifies the expression of a Lagrangian $L_{\rho_0}$ of the type
\begin{equation}\label{unredLagr}
L_{\rho_0}(\mathbf{X},\dot{\mathbf{X}},\mathcal{Q},\dot{\mathcal{Q}})=\ell(\rho,\boldsymbol{u},Q,D_t Q)
\,,
\end{equation}
where $\ell(\rho,\boldsymbol{u},Q,D_t Q)$ is the Lagrangian appearing in the Eulerian action principle \eqref{QStextActPrincFluid}, which has now been written in \eqref{ActPrinc2} in terms of purely Lagrangian variables. {\color{ black}The relation \eqref{unredLagr} means that the Lagrangian description (arising from $L_{\rho_0}$) is equivalent to the Eulerian description (arising from $\ell$) -- in fluid mechanics, this is typically known as `relabelling symmetry'.}

At this point, without entering into the difficult question of computing the Euler-Lagrange equations for $\mathbf{X}$ and $\mathcal{Q}$, we observe that the previous Eulerian variations arise naturally from the definitions above. Thus, one has
\[
\delta\boldsymbol{u}=\delta (\dot{\mathbf{X}}\circ\mathbf{X}^{-1})
\,,\quad
\delta Q= \delta(\mathcal{Q}\circ\mathbf{X}^{-1})
\,,\quad
\delta \rho=\delta (\mathbf{X}_*\rho_0)
\]
along with the relations $\boldsymbol{w}=(\delta\mathbf{X})\circ\mathbf{X}^{-1}$, $\Theta=(\delta\mathcal{Q})\circ\mathbf{X}^{-1}$ and $(\partial_t\mathcal{Q})\circ\mathbf{X}^{-1}=D_t Q$.

Now that we have unfolded the relabelling features of the QS model, it is easy to proceed analogously to the previous section by making the evolution ansatz \eqref{evolution} for biaxial liquid crystal flows. Indeed, with the definitions above, we write
\begin{equation}\label{evol2}
\mathcal{Q}=\mathcal{R}\mathcal{Q}_0\mathcal{R}^{-1}
\end{equation}
(where $\mathcal{R}(\mathbf{x}_0,t)$ is a rotation matrix)
and introduce the angular frequency matrix $\widehat{\omega}=(\partial_t\mathcal{R})\mathcal{R}^{-1}$. Then, the variations  $\delta\mathcal{Q}$ and $\delta\widehat{\omega}$  are expressed as in \eqref{vars1} (upon replacing $\widehat{\nu}$ by $\widehat{\omega}$ and $Q$ by $\mathcal{Q}$). However, in order to obtain the corresponding Eulerian description, we define
\begin{align*}
Q&=(\mathcal{R}\mathcal{Q}_0\mathcal{R}^{-1})\circ\mathbf{X}^{-1}=\mathcal{Q}\circ\mathbf{X}^{-1}
\\
\widehat{\nu}&=(\partial_t\mathcal{R}\,\mathcal{R}^{-1})\circ\mathbf{X}^{-1}=\widehat{\omega}\circ\mathbf{X}^{-1}
\,,
\end{align*}
so that we find
\begin{align*}
&D_t Q=[\widehat{\nu},Q]
\\
&\delta Q=[\widehat{\eta},Q]-(\boldsymbol{w}\cdot\nabla)Q
\\
&\delta\widehat{\nu}
	=\partial_t\widehat{\eta}-(\boldsymbol{w}\cdot\nabla)\widehat{\nu}+(\boldsymbol{u}\cdot\nabla)\widehat{\eta}+[\widehat{\eta},\widehat{\nu}]
\end{align*}
with $\widehat{\eta}=(\delta\mathcal{R}\, \mathcal{R}^{-1} )\circ\mathbf{X}^{-1}$ arbitrary and vanishing at the endpoints. Then, Hamilton's principle associated to the action \eqref{ActPrinc2} is transformed into its Eulerian formulation as
\begin{multline}\label{actprinc3}
\delta\!\int_{t_1}^{t_2}\!\!\!\int\! \bigg(\frac12\rho\|\boldsymbol{u}\|^2+\frac{J}{2}\rho\|[\widehat{\nu},Q]\|^2
\\
-\rho\,\mathcal{U}(\rho)
-\mathcal{F}(Q,\nabla Q)\bigg)\,\de^3{x}\,\de t=0
\end{multline}
and by using the variational relations above, one is led to the Euler-Poincar\'e equations \eqref{QS1}, \eqref{QS3} and
\begin{align}\label{VK}
&	\rho JD_t\big[[\widehat{\nu},Q],Q\big]
	=\big[h,Q\big]
	\,,\qquad\quad\, \  
	D_t Q=[\widehat{\nu},Q]
\end{align}
Again, these equations coincide with the conservative limit of the VK model in \cite{VK1981}, upon setting $J=0$ in order to neglect inertial effects, and they do not involve Lagrange multipliers since the invariants $\Tr Q^n$ are naturally preserved by the evolution of $Q$. In a similar fashion to that of the previous section, the $Q-$equations combine into
\begin{equation}
	[\rho JD_t^2Q-h,Q]
	=0
	\,,
\label{QS-VK}
\end{equation}
which is the restriction of equation \eqref{QS2} under the assumption \eqref{evolution} of rotational evolution.

In conclusion, we have proved that the conservative VK theory of biaxial liquid crystals is a specialization of the QS model under the assumption \eqref{evolution} of purely rotational dynamics, and that this is due to their common action principle \eqref{QStextActPrincFluid}. 

A micropolar fluid version of the above VK equations can be obtained by following the same steps as in the previous section, upon defining 
$
\widehat{\xi}_i:=-\partial_i \mathcal{R}\mathcal{R}^{-1}+\mathcal{R}\,\widehat{\xi}_{0i} \mathcal{R}^{-1}
$ and $\widehat{\gamma}$ such that
\[
\rho\,\widehat{\gamma}:=\int\!\rho_0(\mathbf{x}_0)\widehat{\xi}(\mathbf{x}_0)\,\delta(\mathbf{x}-\mathbf{X}(\mathbf{x}_0,t))\,\de^3 x_0\,.
\]
The equations of motion can be written upon computing
\begin{align*}
	\dot{\widehat{\gamma}}_i+(\boldsymbol{u}\cdot\nabla)\widehat{\gamma}_i+(\partial_i u^k)\widehat{\gamma}_k
	&=[\widehat{\nu},\widehat{\gamma}_i]-\partial_i\widehat{\nu}
	\\
	\delta{\widehat{\gamma}}_i+(\boldsymbol{w}\cdot\nabla)\widehat{\gamma}_i+(\partial_i w^k)\widehat{\gamma}_k
	&=[\widehat{\eta},\widehat{\gamma}_i]-\partial_i\widehat{\eta}
\end{align*}
and by writing $\mathcal{F}(Q,[Q,\widehat{\gamma}])=\Psi(Q,\widehat\gamma)$. Then, upon denoting $p=\rho^2\mathcal{U}'-\Psi$, the action principle \eqref{actprinc3} yields
\begin{align}\label{micro1}
	&\rho D_t {u}_i=-\partial_i
p-\partial_l\!\Tr\!\left(\widehat{\gamma}_i\,\frac{\partial\Psi}{\partial\widehat{\gamma}_l}\right)
\,,\qquad  
D_t Q=[\widehat{\nu},Q]
	\\\label{micro2}
		&\rho JD_t\big[[\widehat{\nu},Q],Q\big]
	=\left[Q,\frac{\partial\Psi}{\partial Q}\right]
	+\left[\frac{\partial\Psi}{\partial\widehat{\gamma}_i},\widehat{\gamma}_i\right]
	-\partial_i\frac{\partial\Psi}{\partial\widehat{\gamma}_i},
	\\\label{micro3}
	&
	D_t{\widehat{\gamma}}_i+(\partial_i u^k)\widehat{\gamma}_k
	=[\widehat{\nu},\widehat{\gamma}_i]-\partial_i\widehat{\nu}
\end{align}
These equations represent the micropolar version of the VK model for flowing liquid crystals. 
\section{Rayleigh dissipation and rotational dynamics }
This section introduces dissipation in the previous liquid crystal models. Although dissipation can be introduced in the Hamiltonian framework by the use of symmetric brackets \cite{delaware1994thermodynamics}, the present framework  makes the use of the Rayleigh dissipation function $\mathscr{R}(Q,\dot{Q})$, in which case the variational principle associated to the QS equation \eqref{QStexture} for texture dynamics {\color{ black}becomes}
\begin{multline}
\delta\!\int_{t_1}^{t_2}\!\!\!\int\! \left(\frac{J}{2}\|\dot Q\|^2
-\mathcal{F}(Q,\nabla Q)
	+\lambda_0\Tr Q\right)\de^3{x}\,\de t
	\\
	=\int_{t_1}^{t_2}\!\!\!\int\Tr\!\left(\frac{\delta \mathscr{R}}{\delta \dot{Q}}\, \delta Q\right)
	\,\de^3{x}\,\de t
	\label{QStextActPrincDis}
\end{multline}
so that \eqref{QStexture} {\color{ black}changes to}
\[
	J\pa_t^2{Q}-h
	=\lambda_0 \boldsymbol{1}-\frac{\delta \mathscr{R}}{\delta \dot{Q}}
	\,.
\]
Here we have used the notation for the functional derivative, whose definition reads
\[
\delta \mathbf{F}(\mathbf{w})=\!\int\frac{\delta \mathbf{F}}{\delta \mathbf{w}}\cdot\delta \mathbf{w}\,\de^3 x
\]

On the other hand, under the rotational evolution \eqref{evolution} for biaxial nematics, one has $\mathscr{R}(Q,\dot{Q})=r(\widehat{\nu},Q)$ with the variations \eqref{vars1}, so that the relation $\delta\mathscr{R}=\delta r$ yields
\[
\frac{\delta r}{\delta \widehat{\nu}}=\left[\frac{\delta \mathscr{R}}{\delta \dot{Q}},Q\right].
\]
Eventually, as noticed already in \cite{GBRaTr2012}, the Euler-Poincar\'e variational principle with dissipation
\begin{multline}
\delta\!\int_{t_1}^{t_2}\!\!\!\int\! \left(\frac{J}{2}\|[\widehat{\nu},Q]\|^2
-\mathcal{F}(Q,\nabla Q)	\right)
	\de^3{x}\,\de t
	\\
	=\int_{t_1}^{t_2}\!\!\!\int\Tr\!\left(\frac{\delta r}{\delta \widehat{\nu}}\, \widehat{\eta}\right)
	\,\de^3{x}\,\de t
\end{multline}
yields the equations of motion
\[
	J\frac{\pa}{\pa t}\big[[\widehat{\nu},Q],Q\big]
	=\big[h,Q\big]{-\frac{\delta r}{\delta \widehat{\nu}}}
	\,,\qquad\ \ 
	\frac{\pa Q}{\pa t}=[\widehat{\nu},Q]
	\,,
\]
which represent the dissipative variant of the conservative VK equations \eqref{VKdyn} for texture dynamics. Again, these combine into
\[
\left[J\ddot{Q}-h+\frac{\delta \mathscr{R}}{\delta \dot{Q}},Q\right]
=0\,,
\]
where the dot notation stands for partial time derivative. The same approach applies to the micropolar theory \eqref{ErText1}-\eqref{ErText2}, whose dissipative version is obtained by inserting the term $-\delta r/\delta \widehat\nu$ into the right hand side of \eqref{ErText1}.

In the case of flowing liquid crystals, dissipation can be introduced by mimicking the same steps as above. For a consistent theory, it is essential to start with the Lagrangian $L_{\rho_0}(\mathbf{X},\dot{\mathbf{X}},\mathcal{Q},\dot{\mathcal{Q}})$ defined in \eqref{unredLagr} and write its dissipative action principle
\begin{multline}
\delta\!\int_{t_1}^{t_2}\!\!\!\int\! L_{\rho_0}(\mathbf{X},\dot{\mathbf{X}},\mathcal{Q},\dot{\mathcal{Q}})\,\de t
	\\
	=\int_{t_1}^{t_2}\!\!\!\int\!\left(\frac{\delta \mathscr{R}_{\rho_0}}{\delta \dot{\mathbf{X}}}\cdot \delta\mathbf{X}+\Tr\!\left(\frac{\delta \mathscr{R}_{\rho_0}}{\delta \dot{\mathcal{Q}}}\, \delta \mathcal{Q}\right)\right)
	\de^3{x}\,\de t
\end{multline}
where $\mathscr{R}_{\rho_0}=\mathscr{R}_{\rho_0}(\mathbf{X},\dot{\mathbf{X}},\mathcal{Q},\dot{\mathcal{Q}})$. Then, upon assuming that $\mathscr{R}_{\rho_0}$ depends on $\mathbf{X}$ only through $\rho=\mathbf{X}_*\rho_0$ and using the same notation and definitions as in the previous section, we can write
\[
\mathscr{R}_{\rho_0}(\mathbf{X},\dot{\mathbf{X}},\mathcal{Q},\dot{\mathcal{Q}})=
\! r(\boldsymbol{u},\rho,Q,D_t{Q})
\]
with 
\begin{align}\label{Raydissvar1}
\frac{\delta r}{\delta \boldsymbol{u}}&=\int\!\frac{\delta \mathscr{R}_{\rho_0}}{\delta \dot{\mathbf{X}}}(\mathbf{x}_0,t)\,\delta(\mathbf{x}-\mathbf{X}(\mathbf{x}_0,t))\,\de^3 x_0
\,,\\\label{Raydissvar2}
\frac{\delta r}{\delta (D_t Q)}&=\int\!\frac{\delta \mathscr{R}_{\rho_0}}{\delta \dot{\mathcal{Q}}}(\mathbf{x}_0,t)\,\delta(\mathbf{x}-\mathbf{X}(\mathbf{x}_0,t))\,\de^3 x_0
\,.
\end{align}
Eventually, 
\begin{multline*}
\int\!\left(\frac{\delta \mathscr{R}_{\rho_0}}{\delta \dot{\mathbf{X}}}\cdot \delta\mathbf{X}+\Tr\!\left(\frac{\delta \mathscr{R}_{\rho_0}}{\delta \dot{\mathcal{Q}}}\, \delta \mathcal{Q}\right)\right)
	\de^3{x}
	\\
	=
	\int\!\left(\frac{\delta r}{\delta \boldsymbol{u}}\cdot \boldsymbol{w}+\Tr\!\left(\frac{\delta r}{\delta (D_t Q)}\, \delta {Q}\right)\right)
	\de^3{x}\,,
\end{multline*}
{\color{ black}which then replaces the right hand side of \eqref{QStextActPrincFluid} so that} the QS equations  \eqref{QS1}-\eqref{QS3} become
\begin{align}\label{QS1diss}
&\rho D_t \boldsymbol{u}= -\nabla
p-\partial_l\!\left(\frac{\partial
\mathcal{F}}{\partial{Q}_{ij\,,l}}\nabla{Q}_{ij}\right)-\frac{\delta r}{\delta \boldsymbol{u}},
\\ \label{QS2diss}
&\rho JD_t^2{Q}-h=\lambda_0\boldsymbol{1}-\frac{\delta r}{\delta {(D_tQ)}},
\\ \label{QS3diss}
&
D_t \rho+\rho\operatorname{div}\boldsymbol{u}=0
\,.
\end{align}
Alternatively, one may proceed by assuming the evolution \eqref{evol2} in $\mathscr{R}_{\rho_0}$ to write 
\[
\int \!\mathscr{R}_{\rho_0}(\mathbf{X},\dot{\mathbf{X}},\mathcal{Q},\dot{\mathcal{Q}})\,\de^3 x_0=\int
\! r(\boldsymbol{u},\rho,\widehat{\nu},Q)\,\de^3 x\,,
\]
with \eqref{Raydissvar1} and
\begin{align*}
\frac{\delta r}{\delta \widehat{\nu}}
&=\int\!\left[{\frac{\delta \mathscr{R}}{\delta \dot{\mathcal{Q}}}(\mathbf{x}_0,t),\mathcal{Q}(\mathbf{x}_0,t)}\right]\delta(\mathbf{x}-\mathbf{X}(\mathbf{x}_0,t))\,\de^3 x_0
\,.
\end{align*}
Then, 
\begin{multline*}
\int\!\left(\frac{\delta \mathscr{R}_{\rho_0}}{\delta \dot{\mathbf{X}}}\cdot \delta\mathbf{X}+\Tr\!\left(\frac{\delta \mathscr{R}_{\rho_0}}{\delta \dot{\mathcal{Q}}}\, \delta \mathcal{Q}\right)\right)\delta	\de^3{x}
	\\
	=
	\int\!\left(\frac{\delta r}{\delta \boldsymbol{u}}\cdot \boldsymbol{w}+\Tr\!\left(\frac{\delta r}{\delta \widehat{\nu}}\, \widehat{\eta}\right)\right)
	\de^3{x}
\end{multline*}
and the dissipative version of the VK equations is obtained by inserting the term $-\delta r/\delta \widehat\nu$ into the right hand side of the first of \eqref{VK}, which is then accompanied by \eqref{QS1diss} and \eqref{QS3diss}. After verifying that ${\delta r}/{\delta \widehat{\nu}}=[{\delta r}/{\delta (D_t Q)},Q]$, these equations combine into 
\begin{align*}
&\rho D_t \boldsymbol{u}= -\nabla
p-\partial_l\!\left(\frac{\partial
\mathcal{F}}{\partial{Q}_{ij\,,l}}\nabla{Q}_{ij}\right)-\frac{\delta r}{\delta \boldsymbol{u}},
\\ 
&\left[\rho JD_t^2{Q}-h+\frac{\delta r}{\delta (D_t Q)},Q\right]
=0,
\\ 
&
D_t \rho+\rho\operatorname{div}\boldsymbol{u}=0
\,.
\end{align*}

At this point, the micropolar formulation of VK dynamics can also be presented in dissipative form by following precisely the same steps as above. This consists of adding the term $-\delta r/\delta \boldsymbol{u}$ into the right hand side of the first in \eqref{micro1} and the term $-\delta r/\delta \widehat\nu$ into the right hand side of \eqref{micro2}.

It is important to emphasize that different Rayleigh functions produce different relaxation dynamics and therefore different physical results. Here, we shall not dwell upon the difficult question of what the correct expression of the Rayleigh function should be; see \cite{sonnet2004continuum}. Still, the results presented here remain valid independently of the particular choice of Rayleigh functional.

\section{Scaling and eigenvalue dynamics}

The evolution \eqref{evolution} is isospectral -- that is, the eigenvalues of $Q$ are preserved. However, more general cases are also allowed in liquid crystal dynamics. Then, the purely rotational motion can be extended in order to incorporate nontrivial eigenvalue dynamics. In the present geometric context this is achieved by allowing the evolution law to admit scaling. In light of this, one choice is to let $Q$ evolve under dilations as well as rotations, that is
\begin{equation}\label{confev}
	Q=\lambda^2RQ_0R^T,
\end{equation}
where $\lambda(\bx,t)$ is a time-dependent positive function. For example, in the case of an initial uniaxial phase $Q_0=\sigma\left(\mathbf{n}_0\mathbf{n}_0-\boldsymbol{1}/3\right)$ (for some constant $\sigma$), the above relation yields
\[
	Q=\sigma\lambda^2\!\left(\mathbf{n}\mathbf{n}-\frac13\boldsymbol{1}\right)
\!,
\]
where $\sigma\lambda^2$ may now acquire the meaning of a scalar order parameter. A more general situation can be studied by introducing two conformation tensors $A$ and $B$ such that $Q=\lambda^2RAR^T+\zeta^2R B R^T$. However, this possibility is left for future work.

The evolution law \eqref{confev} comes out in a natural fashion by considering the congruence transformation $Q=g Q_0 g^T$ for some invertible matrix $g(\bx,t)$. {\color{ black}Indeed, as shown in  Appendix \ref{Appendix}, a natural choice is a matrix $g$ of the form $g=\lambda R$}. As an additional remark, we recall that such conformal rotations play a crucial role in Eringen's description of microstretch fluids \cite{eringen2001microcontinuum1,eringen2001microcontinuum2}, as enlightened in \cite{gay2009geometric}. 
This motivates us to investigate  the interplay between scaling and rotational degrees of freedom. To this purpose, we simply extend the Euler-Poincar\'e theory from previous sections to accommodate dilations.

Upon using the evolution law \eqref{confev}, one follows the procedure outlined in Section \ref{sec:flows} by defining 
\begin{align*}
Q&=\mathcal{Q}\circ\mathbf{X}^{-1}=(\lambda^2\mathcal{R}\mathcal{Q}_0\mathcal{R}^{-1})\circ\mathbf{X}^{-1},
\\
\widehat{\nu}&=(\partial_t\mathcal{R}\,\mathcal{R}^{-1})\circ\mathbf{X}^{-1}
\,,
\\
\Lambda&=(\lambda^{-1}\partial_t\lambda)\circ\mathbf{X}^{-1}
\,,
\end{align*}
so that the variational principle \eqref{actprinc3} is modified to
\begin{multline}
\delta\!\int_{t_1}^{t_2}\!\!\!\int\! \bigg(\frac12\rho\|\boldsymbol{u}\|^2+\frac{J}{2}\rho\|[\nu,Q]\|^2+2\rho J\Lambda^2\|Q\|^2
\\
-\rho\,\mathcal{U}(\rho)
-\mathcal{F}(Q,\nabla Q)
	\bigg)\,\de^3{x}\,\de t=0\,,
	\label{}
\end{multline}
with the  variations
\begin{align}\label{vars2}
&\delta Q=[\widehat{\eta},Q]+2\Xi Q-(\boldsymbol{w}\cdot\nabla)Q
\\
&\delta\widehat{\nu}
	=\partial_t\widehat{\eta}-(\boldsymbol{w}\cdot\nabla)\widehat{\nu}+(\boldsymbol{u}\cdot\nabla)\widehat{\eta}+[\widehat{\eta},\widehat{\nu}]
\,,
\\
&\delta\Lambda=\partial_t\Xi-(\boldsymbol{w}\cdot\nabla)\Xi+(\boldsymbol{u}\cdot\nabla)\Lambda\,,
\end{align}
where  $\widehat{\eta}$, $\Xi$ and $\boldsymbol{w}$ are arbitrary and vanish at the endpoints. Then, the resulting Euler-Poincar\'e equations extend the VK model to incorporate scaling:
\begin{align}\label{QS1bis}
&\rho D_t \boldsymbol{u}= -\nabla
p-\partial_l\!\left(\frac{\partial
\mathcal{F}}{\partial{Q}_{ij\,,l}}\nabla{Q}_{ij}\right),
\quad
D_t \rho=-\rho\operatorname{div}\boldsymbol{u},
\\ \label{QS2bis}
&\rho JD_t\big[[\widehat{\nu},Q],Q\big]
	=\big[h,Q\big]
,
\quad\ 
	D_t Q=[\widehat{\nu},Q]+2\Lambda Q,
\\ \label{QS3bis}
&
2\rho J \|Q\|^2(D_t\Lambda+2\Lambda^2)=\rho J \|[\widehat\nu,Q]\|^2+\operatorname{Tr}(h Q)
\,,
\end{align}
where we observe that the first three equations remain unchanged, while the modified $Q-$tensor dynamics produces an additional equation for the scaling quantity $\Lambda$ (the latter quantifies the time rate of $\operatorname{Tr}Q^n$ far all integers $n$). 
 Also, notice that the dynamics above can be rewritten upon replacing \eqref{QS2bis} and \eqref{QS3bis} by
\[
[\rho JD_t^2Q-h,Q]
	=0
\,,\qquad\quad 
\operatorname{Tr}\!\big((\rho J D_t^2Q-h)Q\big)=0
\,,
\]
where the latter is obtained by combining \eqref{QS3bis} with \eqref{QS2bis}. Note that the above dynamical system is also a special case of the QS model,  since the former is obtained by making use of the ansatz \eqref{confev} in the variational principle \eqref{QStextActPrincFluid} underlying the QS equations of motion \eqref{QS1}-\eqref{QS3}.

At this stage, inserting dissipation is straightforward as it can be done by simply applying the method outlined in the previous section. Therefore, we shall just state the final results, which are as follows:
\begin{align*}
&\rho D_t \boldsymbol{u}= -\nabla
p-\partial_l\!\left(\frac{\partial
\mathcal{F}}{\partial{Q}_{ij\,,l}}\nabla{Q}_{ij}\right)-\frac{\delta r}{\delta \boldsymbol{u}},
\\ 
&\left[\rho JD_t^2{Q}-h+\frac{\delta r}{\delta (D_t Q)},Q\right]
=0,
\\
&
J\rho\operatorname{Tr}(QD_t^2Q)=\operatorname{Tr}\!\left(\!\left(h-\frac{\delta r}{\delta (D_t Q)}\right)\!Q\right),
\\ 
&
D_t \rho+\rho\operatorname{div}\boldsymbol{u}=0
\,.
\end{align*}
As before, the Rayleigh dissipation function is left arbitrary to allow for different modeling options. Possible choices of Rayleigh function are presented in \cite{sonnet2004continuum}.

\section{Conclusions and open questions}
This paper has compared three apparently different approaches to $Q-$tensor dynamics in the theory of liquid crystals. Starting from simple texture dynamics, the VK theory {\color{ black}for biaxial nematics was shown to emerge from the QS model}
under the preliminary assumption of conservative dynamics and by retaining inertial effects.  In addition, a micropolar variant of the VK theory has been formulated by following Eringen's definition of the wryness tensor \cite{Eringen1997}. Already at this stage the use of Euler-Poincar\'e reduction theory is advantageous, and even more so in the case of flowing liquid crystals, which was the subject of Section 3. The relation between QS theory and VK dynamics was shown to hold in this more general case without essential modifications and the same holds for the corresponding micropolar variant. Finally, we applied Euler-Poincar\'e reduction to the dissipative action principle, thereby showing how rotational dynamics \eqref{evolution} can also be taken into account for dissipative liquid crystal flows. Notice that all the results obtained in this paper also apply to the special case of incompressible flows.

Another natural question was addressed regarding the possibility of $Q-$tensor flows that are more general than \eqref{evolution}. Indeed, while an isospectral flow of $Q$ is justified for biaxial molecules, other types of nematic molecules require more general evolutions which are then influenced by the terms $\Tr Q^n$ in the Landau-de Gennes free energy. A possible strategy to tackle this question was pursued by making use of the conformal rotation group, so that \eqref{evolution} is replaced by \eqref{confev}. This approach  was followed by Eringen in his theory of microstretch fluids \cite{eringen2001microcontinuum1,eringen2001microcontinuum2}, and its geometric features have been exploited in \cite{gay2009geometric}.  In the present paper, a full dynamical theory was developed based on the Qian-Sheng Lagrangian and  dissipation was also included.

The work in this paper follows in the same direction as recent work \cite{GBRaTr2012}, which showed how Eringen's micropolar theory includes Ericksen-Leslie dynamics as a special case. It is hoped that the identification of relations between different liquid crystal theories will help provide a better understanding of the essential features captured in the various models. As mentioned in the introduction, an outstanding question concerns possible deep relations between the Harvard theory and EL dynamics. The answer to this question is among the most difficult, since the order parameter possesses essentially different transformation properties in the two models.

\paragraph{Acknowledgements} The authors are indebted to Paul Skerritt for a crucial step in the proof reported in the Appendix \ref{Appendix}. Also, the authors benefited from stimulating conversations with Fran\c{c}ois Gay-Balmaz and Arghir Dani Zarnescu. {\color{ black}Part of this work was developed at the Isaac Newton Institute for Mathematical Sciences (Cambridge, UK), whose hospitality is greatly acknowledged. Financial support by the Leverhulme Trust Research Project Grant No. 2014-112 is also acknowledged.}

\appendix

\section{Dilations, rotations and traces\label{Appendix}}

{\color{ black}
This appendix shows how conformal rotations may arise from the transformation properties of traceless symmetric matrices. In particular, we shall show the following
\begin{lemma}
Let $g\in\mathrm{GL}(n,\mathbb{R})$ be such that
\[
gQg^T\in\mathrm{sym}_0(3)\,,\quad\forall Q\in\mathrm{sym}_0(3),
\]
where $\mathrm{sym}_0(3)$ denotes the space of traceless symmetric matrices. 
Then, $g$ is of the fom
\[
g=\lambda R
\]
with $\lambda\in\mathbb{R}^+$ and $R\in\mathrm{SO}(3)$.
\end{lemma} 
\paragraph{Proof.}
Clearly the quantity $gQg^T\!\!,\ g\in$ GL(3) is symmetric. To preserve tracelessness one has $\big<g^Tg,Q\big>=0$ as an inner product on sym(3). Since this form is nondegenerate, $g^Tg$ must be an element of sym$^{\bot}_0$(3), the orthogonal complement to the traceless subspace of sym(3). Then noting the following:
\begin{align*}
	&\mathrm{dim\ sym}(3)=6\\
	&\mathrm{dim\ sym}_0(3)=5\\
	&\mathrm{dim\ sym^{\bot}_0}(3)=1,
\end{align*}
inferred by counting, the only symmetric and tracefull 1-dimensional space orthogonal to every element of sym$_0$(3) is $\lambda^2\boldsymbol{1}$ for $\lambda^2\in\mathbb{R}$, since $g^Tg$ is positive definite. Therefore $g=\lambda R$ for $R\in$ SO(3).
\quad $\blacksquare$}


\end{multicols}
\end{document}